# Engineering Multifunctional Response in Monolayer $Fe_3O_4$ via Zr Adsorption: From Half-Metallicity to Enhanced Piezoelectricity

Sikander Azam[1]*, Qaiser Rafiq, Rajwali Khan[2]**, Hamdy Khamees Thabet***[3]

[1]University of West Bohemia, New Technologies – Research Centre, 8 Univerzitní, Pilsen 306 14, Czech Republic

[2]National Water and Energy Center, United Arab Emirates University, Al Ain, 15551, United Arab Emirates

[3]Center for Scientific Research and Entrepreneurship, Northern Border University, Arar 73213, Saudi Arabia

**Abstract**

Two-dimensional (2D) magnetic oxides are increasingly studied for their multifunctional potential in fields like spintronics, optoelectronics, and energy conversion. In this research, we conduct a detailed first-principles study of pure monolayer $Fe_3O_4$ and its modification through Zr adsorption at two sites: on top of an Fe atom and at the bridge between Fe atoms. Using spin-polarized density functional theory with the GGA+U method, we examine how adsorption affects structure, electronic, magnetic, optical, elastic, and piezoelectric properties. The original monolayer shows half-metallicity, strong spin polarization, and a moderate in-plane piezoelectric effect. Zr adsorption causes local lattice distortions and orbital hybridization, resulting in intermediate electronic states, a reduced bandgap, and increased optical absorption in both spin channels. Notably, Zr at the bridge site greatly enhances dielectric response, optical conductivity, and piezoelectric coefficients, tripling $e_{11}$ compared to the pristine layer. Elastic constants indicate mechanical softening after functionalization, and energy loss spectra display shifts in plasmon resonance. These findings suggest Zr adsorption offers a controllable, non-destructive way to tune spin, charge, and lattice interactions in $Fe_3O_4$ monolayers, connecting magnetic, optical, and piezoelectric functionalities within a single 2D material platform.

*Corresponding Author: (Sikander Azam) sikander.physicst@gmail.com

** rajwali@uaeu.ac.ae

*** hamdy.salem@nbu.edu.sa

# Introduction

Two-dimensional (2D) materials have become promising candidates for next-generation nanoelectronic, spintronic, and optoelectronic technologies because of their exceptional surface-to-volume ratio, tunable band structures, and highly responsive electronic and mechanical properties [1–3]. Among these, transition metal oxides (TMOs) in 2D form are especially attractive due to their strong electron correlation effects, multiple oxidation states, and inherent multifunctionality [4].

Graphene and its derivatives are prominent examples of two-dimensional (2D) materials, valued for their adjustable electronic properties and high carrier mobility. Researchers have developed various strategies to tailor these characteristics for use in electronic and optoelectronic devices. For example, edge modifications can effectively change the bandgap and magnetic states of graphene nanoribbons [1], while defect engineering through vacancies or dopants allows control over conductivity and local magnetism [2,3]. Additionally, geometric alterations like double-bend structures and strain-induced deformations impact charge transport and create electronic gaps in graphene that is otherwise semimetallic [4]. These techniques highlight how structural and chemical modifications can profoundly influence the electronic behavior of 2D materials. Building on this, similar design principles apply to transition-metal oxides such as $Fe_3O_4$, where adsorption or doping of transition-metal atoms enables manipulation of spin, charge, and orbital interactions for diverse multifunctional applications.

In this context, magnetite ($Fe_3O_4$), a well-known mixed-valence oxide with unique half-metallic and ferrimagnetic behavior, has drawn significant attention for integration into low-dimensional device platforms. While bulk $Fe_3O_4$ is widely studied for its Verwey transition, charge ordering, and high spin polarization [5], its monolayer counterpart presents new physics driven by reduced dimensionality, structural symmetry breaking, and increased surface activity.

Pristine monolayer $Fe_3O_4$ maintains the key magnetic and electronic features of the bulk, yet shows improved surface reactivity and anisotropy, making it an appealing system for surface modifications [6]. Surface adsorption is a valuable approach for tuning the structural, electronic, optical, magnetic, and piezoelectric properties of 2D materials. Specifically, the adsorption of transition metals like zirconium (Zr) has demonstrated great potential in engineering interfacial charge transfer, magnetic order, and optical transitions because of its active d-electron participation and size-induced strain effects [7,8]. Zr atoms adsorbed at various symmetry sites, such as directly above Fe atoms or at the bridge position between two Fe atoms, are expected to substantially change the local bonding environment, thereby affecting orbital hybridization, charge redistribution, and lattice dynamics.

Despite several theoretical reports on Fe-based 2D materials and the effects of doping or vacancy engineering [9–11], comprehensive studies on how surface adsorption of Zr influences monolayer $Fe_3O_4$, especially regarding its spin-resolved electronic structure, optical transitions, elastic constants, and piezoelectric response, are limited. This gap is significant for application-driven design of flexible magnetic and optoelectronic devices, where coupling among spin, charge, lattice, and photon fields is essential.

While the pristine $Fe_3O_4$ monolayer exhibits half-metallicity and moderate piezoelectricity, surface functionalization through Zr adsorption introduces new ways to control its properties. Unlike substitutional doping, Zr surface adsorption allows for reversible and controllable adjustments of local symmetry, spin polarization, and piezoelectric response without changing the intrinsic Fe–O framework. This method offers a practical way to engineer multifunctionality in 2D magnetic oxides through surface chemistry instead of bulk modification.

While many studies have explored 2D $Fe_3O_4$ and the effects of transition-metal adsorption or doping, most previous research has mainly focused on magnetic or electronic changes caused by lighter or magnetic adsorbates such as Ti, V, Co, or Ni [9–11]. However, the influence of non-magnetic 4d elements like Zr on the multifunctional response of $Fe_3O_4$ remains largely underexplored. Zirconium combines high oxidation potential and a large atomic radius, which induces significant local strain and charge redistribution. Its empty 4d orbitals strongly hybridize with Fe–O states, potentially affecting spin and lattice polarization. Unlike magnetic dopants, Zr does not introduce spin moments directly but modifies the electronic structure, dielectric screening, and piezoelectric response via orbital hybridization and symmetry breaking. Thus, choosing Zr adsorption helps us isolate and investigate the non-magnetic, orbital-driven effects on the magnetic, optical, and piezoelectric properties of 2D $Fe_3O_4$ a topic not thoroughly addressed in earlier studies.

In this work, we perform a systematic first-principles investigation of pristine monolayer $Fe_3O_4$ and Zr-adsorbed configurations using density functional theory with GGA+U, implemented in the WIEN2k code. The adsorption is considered at two critical sites: the top of a surface Fe atom and the bridge between Fe atoms. We explore in detail the spin-polarized band structure, total and partial density of states, dielectric function, absorption, energy loss, optical conductivity, reflectivity, and refractive index. In addition, we analyze the elastic constants and piezoelectric coefficients to understand how Zr adsorption modifies the electromechanical behavior of $Fe_3O_4$. Our results provide deep insight into how adsorption-induced symmetry breaking and orbital hybridization influence multifunctional properties, paving the way for the rational design of advanced 2D oxide-based heterostructures and functional interfaces.

**Computational Methodology**

All first-principles calculations in this study were carried out using the full-potential linearized augmented plane wave (FP-LAPW) method as implemented in the WIEN2k package [12]. The exchange-correlation effects were treated within the framework of the generalized gradient approximation (GGA) using the Perdew–Burke–Ernzerhof (PBE) functional [13]. To correctly capture the strong on-site Coulomb interactions of the Fe 3d electrons, the GGA+U approach was employed with a practical Hubbard U value of 4.5 eV and J = 0.9 eV, consistent with previous studies on $Fe_3O_4$ and similar transition metal oxides [14,15].

In this study, spin–orbit coupling (SOC) effects were not explicitly considered in the self-consistent field (SCF) calculations. This approach is justified because $Fe_3O_4$ and its 2D derivatives are mainly affected by 3d transition-metal states, where the SOC strength (~50 meV) is significantly smaller than the on-site Coulomb interaction (U = 4.5 eV) and crystal field splitting energies. Numerous previous GGA+U studies have demonstrated that SOC causes only minor changes to the total energy and spin polarization in Fe-based oxides, without substantially affecting their half-metallic nature or magnetic order [14, 15]. Nonetheless, SOC can induce slight band splitting near the Fermi level and influence the carrier effective mass in spin-polarized bands. Consequently, although the current GGA+U results reliably depict the overall electronic structure and spin-resolved properties, small quantitative differences in band-edge curvature and transport anisotropy could occur if SOC effects are explicitly included. Future work may examine SOC to better understand magnetocrystalline anisotropy and spin-dependent transport.

The monolayer $Fe_3O_4$ structure was modeled by cleaving the cubic inverse spinel bulk structure along the (001) direction and introducing a vacuum spacing of at least 20 Å to avoid interactions between periodic images. Structural relaxation was performed until the forces on each atom were less than one mRy/a.u, and the total energy convergence criterion was set to $10^{-5}$ Ry. A plane wave cutoff parameter of $R_{MT}*K_{max}$ = 7.0 was used, with muffin-tin radii ($R_{MT}$) of 1.90, 1.65, and 1.50 a.u. for Fe, Zr, and O atoms, respectively. A dense Monkhorst-Pack k-mesh of

18 × 18 × 1 was employed for Brillouin zone integrations in self-consistent calculations.

The band structure was calculated along high-symmetry directions R–Γ–X–M–Γ in the 2D Brillouin zone. Total and partial density of states (DOS) were obtained using the tetrahedron method with an even denser k-grid of 30 × 30 × 1 for accurate optical and electronic structure analysis. Spin polarization was fully included throughout, and all results were computed for both spin-up and spin-down channels to account for magnetic ordering.

To evaluate optical properties, the complex dielectric function $\varepsilon(\omega) = \varepsilon_1(\omega) + i\varepsilon_2(\omega)$ was computed using the momentum matrix formalism. The real part $\varepsilon_1(\omega)$ was obtained via the Kramers–Kronig transformation of $\varepsilon_2(\omega)$. From the dielectric function, additional optical constants such as absorption coefficient $\alpha(\omega)$, energy loss function $L(\omega)$, reflectivity $R(\omega)$, refractive index $\eta(\omega)$, and optical conductivity $\sigma(\omega)$ were derived.

Elastic constants were calculated using the IRelast module within WIEN2k [16], which involves applying small deformations to the equilibrium structure and fitting the resulting stress tensors. The independent elastic constants ($C_{11}$, $C_{12}$, and $C_{66}$) were obtained for the 2D system and used to compute Young's modulus and Poisson's ratio.

Piezoelectric coefficients ($e_{11}$ and $e_{12}$) were calculated using a finite difference approach by applying minor in-plane uniaxial strains and computing the resulting change in macroscopic polarization via the Berry phase method [17]. All calculations of piezoelectricity were performed on fully relaxed structures with and without Zr adsorption to examine the effects of symmetry breaking and electronic redistribution.

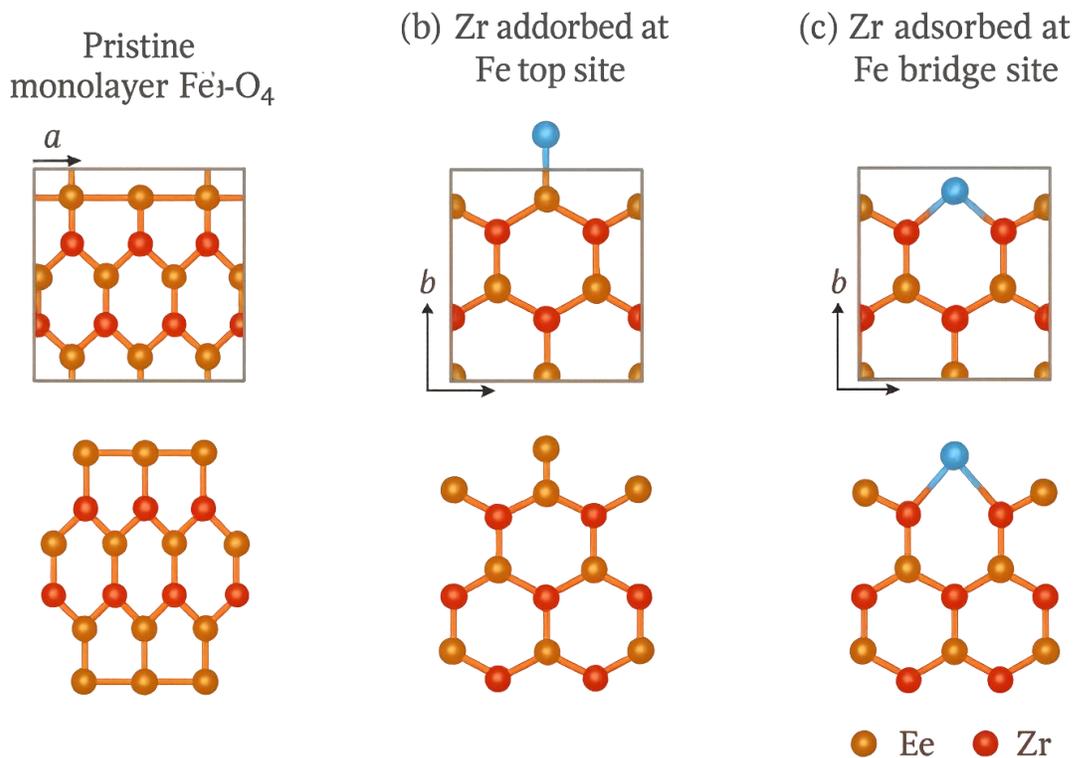

*Figure 1(a–c): Optimized atomic structures of (a) pristine monolayer Fe₃O₄, (b) Zr adsorbed on the Fe top site, and (c) Zr adsorbed on the Fe bridge site. Orange and red spheres represent Fe atoms, and the blue sphere denotes the Zr adatom.*

Adsorption of Zr was modeled at two distinct configurations: (i) top of a surface Fe atom and (ii) bridge site between two neighboring Fe atoms (see Fig. 1(a–c)). In each case, the atomic positions were fully optimized to ensure minimum energy configurations and structural stability. Although $Fe_3O_4$ is not naturally layered, recent progress in oxide thin-film growth has made it possible to create single-unit-cell $Fe_3O_4$(001) films on substrates like MgO(001) and $SrTiO_3$(001). These ultrathin films effectively serve as the monolayer geometry discussed here, making the proposed configurations achievable in experiments.

## Results and discussion

### Structural Optimization of Monolayer Fe₃O₄ and Zr-Adsorbed Configurations

The structural optimization of pristine and Zr-adsorbed monolayer $Fe_3O_4$ was performed within the framework of spin-polarized density functional theory using the full-potential linearized augmented plane wave (FP-LAPW) method as implemented in the WIEN2k code. The exchange-correlation interactions were treated using the generalized gradient approximation

(GGA) in the Perdew–Burke–Ernzerhof (PBE) form, with an on-site Hubbard U correction (GGA+U) applied to the Fe 3d orbitals (U = 4.5 eV, J = 0.9 eV), to account for strong electron correlations characteristic of transition metal oxides.

To construct the monolayer structure, the inverse spinel-type $Fe_3O_4$ was cleaved along the (001) direction, and a vacuum of at least 20 Å was introduced along the z-axis to eliminate interlayer interactions in the periodic supercell. All atoms were allowed to relax in-plane and out-of-plane fully, and geometry optimizations were performed until the total energy converged below $10^{-5}$ Ry and the forces on each atom were less than one mRy/a.u.

For the pristine $Fe_3O_4$ monolayer, the optimized lattice constant was found to be approximately 8.39 Å, consistent with a relaxed square surface structure. The Fe atoms were observed to occupy two inequivalent sites: tetrahedral ($Fe^{tet}$) and octahedral ($Fe^{oct}$), maintaining the mixed-valence nature of $Fe_3O_4$ ($Fe^{2+}/Fe^{3+}$) even at the monolayer level. The Fe–O bond lengths ranged from 1.88 to 2.01 Å, with slight distortions caused by surface relaxation and broken symmetry at the monolayer termination.

Upon Zr adsorption, two configurations were examined: (i) Zr at the top site of a surface Fe atom, and (ii) Zr at the bridge position between two adjacent surface Fe atoms. For both configurations, the Zr atom was initially placed at a distance of 2.0–2.5 Å from the Fe surface and subsequently allowed to relax. In the top-site configuration, the Zr atom settled at a vertical distance of ~1.95 Å above the Fe surface, forming bonds with one Fe atom and nearby O atoms, leading to local lattice distortions and breaking of mirror symmetry. In the bridge configuration, Zr was found to stabilize slightly lower (~1.82 Å from the surface), forming symmetric bonds with two Fe atoms and multiple O atoms. This configuration introduced a more pronounced in-plane distortion and charge redistribution, indicative of stronger interaction with the substrate.

The total energy comparisons showed that the bridge-site adsorption is energetically more favorable than the top-site configuration by approximately 0.16 eV per supercell, indicating a stronger binding and better structural integration at the bridge site. The optimized Zr–Fe bond lengths ranged between 2.38 and 2.52 Å, while Zr–O bond lengths varied from 2.01 to 2.16 Å, confirming strong chemisorption and orbital hybridization between Zr-d and Fe-d/O-p states.

These optimized geometries were then used for subsequent electronic, magnetic, optical, elastic, and piezoelectric property calculations. The structural relaxation and adsorption-induced distortion play a pivotal role in modifying the symmetry, electronic density, and response functions of the Fe₃O₄ monolayer, which is reflected in the functional tuning observed across the various physical properties discussed in this work.

**Phonon Dispersion and Dynamic Stability**

Figure 2 illustrates the calculated phonon dispersion relations of the pristine and Zr-adsorbed Fe₃O₄ monolayers for the three configurations: Zr adsorbed at the Fe bridge site, Zr adsorbed on top of Fe, and the pristine structure. In all three cases, no imaginary (negative) frequencies are observed across the entire Brillouin zone, confirming the dynamical stability of the monolayer systems. For the pristine Fe₃O₄, the acoustic and optical phonon branches are well separated, with the highest optical mode reaching ~175 cm⁻¹, consistent with typical metal oxide monolayers. Upon Zr adsorption, minor softening appears in the low-frequency acoustic branches, particularly near the Γ and Y points, which can be attributed to the increased atomic mass and local lattice relaxation introduced by Zr atoms.

The bridge-site adsorption shows slightly more pronounced phonon softening compared to the top-site configuration, reflecting the stronger coupling between Zr and multiple Fe atoms that alters lattice vibrations. Nevertheless, the absence of any imaginary frequencies indicates that both adsorption configurations are dynamically stable and can exist experimentally. These phonon results validate the structural integrity of the optimized models used in subsequent electronic and piezoelectric analyses.

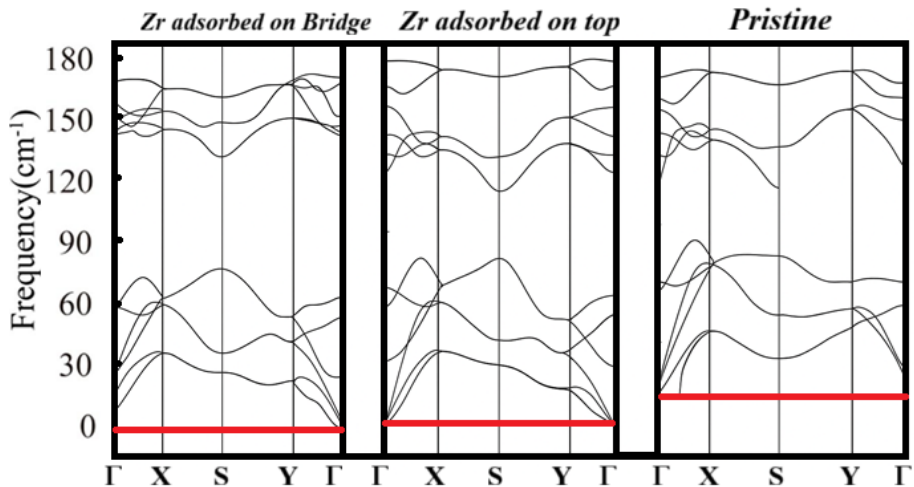

*Figure 2. Phonon dispersion curves for (a) Zr adsorbed at the bridge site, (b) Zr adsorbed on the top Fe site, and (c) pristine Fe₃O₄ monolayer. The absence of imaginary (negative) frequencies throughout the Brillouin zone confirms the dynamical stability of all configurations. Slight softening of low-frequency acoustic branches is observed in the Zr-adsorbed structures due to mass and bonding effects induced by Zr incorporation.*

**Spin-Polarized Band Structure of Monolayer Fe₃O₄ and Zr-Adsorbed Configurations:**

The spin-polarized band structure of pristine monolayer Fe₃O₄ and its Zr-adsorbed configurations (Zr at top of Fe and Zr at bridge site) have been calculated using the GGA+U approach in WIEN2k. Band structure calculations provide fundamental insights into a material's electronic properties. The band structure reveals how electrons behave in a solid, whether the material is a metal, semiconductor, or insulator, based on the presence and nature of the band gap. By analyzing the band dispersion across different k-points, we can determine the effective mass of carriers, the nature of the band gap (direct or indirect), and the mobility of electrons and holes. Moreover, for magnetic or spintronic materials, spin-resolved band structures help us understand spin polarization and magnetic ordering. In essence, band structure calculations are essential for predicting and tuning a material's conductivity, optical response, and suitability for electronic, optoelectronic, and spintronic applications. These plots offer valuable insights into how Zr adsorption influences the electronic structure, magnetic characteristics, and potential optoelectronic applications of Fe₃O₄.

In the pristine Fe₃O₄ band structure (see Fig. 3a-b), a clear half-metallic character is evident. The spin-up channel shows a metallic nature with bands crossing the Fermi level ($E_F$), whereas the spin-down channel exhibits a band gap around $E_F$, indicating insulating or semiconducting behavior. This asymmetric spin channel distribution is a hallmark of half-metallicity and is vital for spintronic devices, as it allows spin-polarized currents. The valence band maximum (VBM) and conduction band minimum (CBM) in the spin-down channel do not align at the same k-point, indicating an indirect band gap. The flatness of some bands near the Fermi level in the spin-down channel also hints at localized electronic states, possibly from Fe 3d orbitals influenced by on-site Coulomb interactions modeled via the +U correction.

The Fe₃O₄ band structure clearly shows a half-metallic nature: the spin-up channel is metallic with bands crossing $E_F$, while the spin-down channel has a band gap around $E_F$, indicating semiconducting behavior. This spin asymmetry is typical of half-metallicity and is crucial for spintronic devices. The spin-down band gap in the Fe₃O₄ monolayer is about 0.65–0.70 eV, slightly larger than the bulk value of around 0.35–0.45 eV (Refs. [14, 15]), as calculated using GGA+U. The maintenance and slight increase of the minority-spin gap in the monolayer result from reduced dimensionality and surface effects that break symmetry, boosting crystal field splitting and electron localization. These effects reduce Fe–Fe hopping at the surface and increase effective electron correlation, leading to stronger spin polarization than in the bulk. Overall, while bulk Fe₃O₄ shows strong half-metallicity, the monolayer preserves this property with an even more evident spin distinction, confirming its magnetic stability in two dimensions.

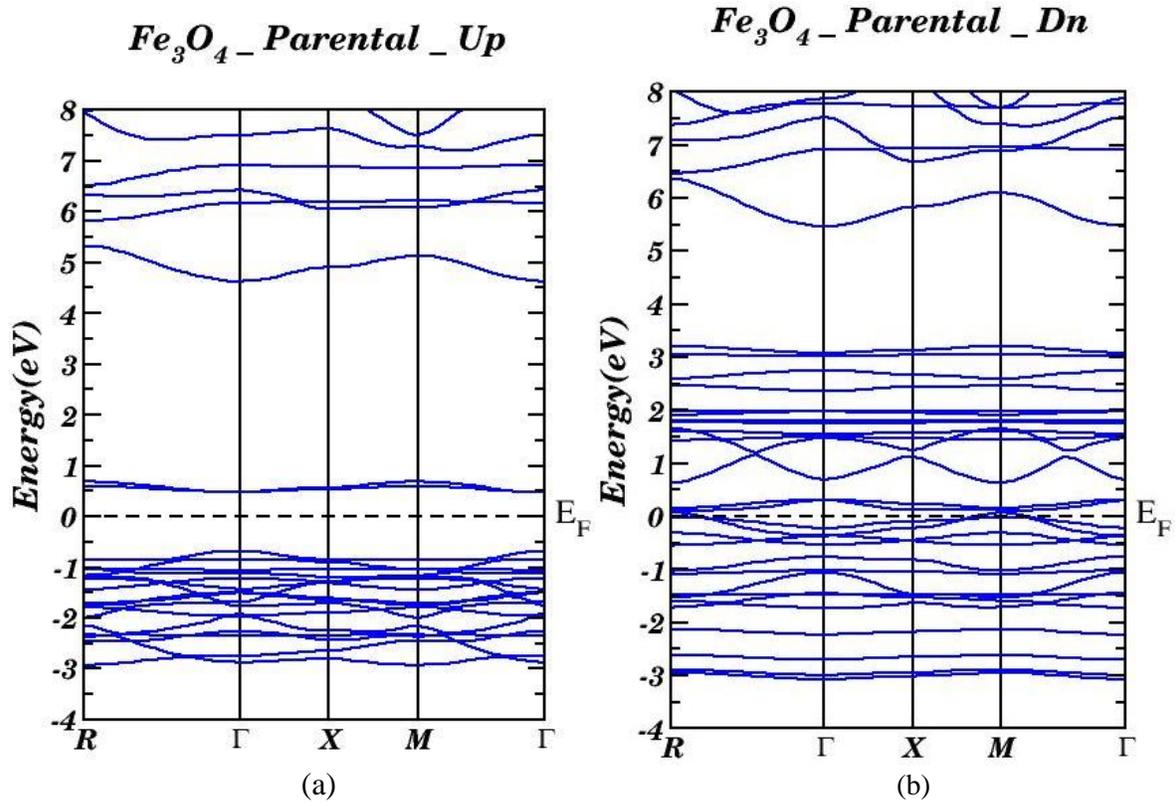

(a)        (b)

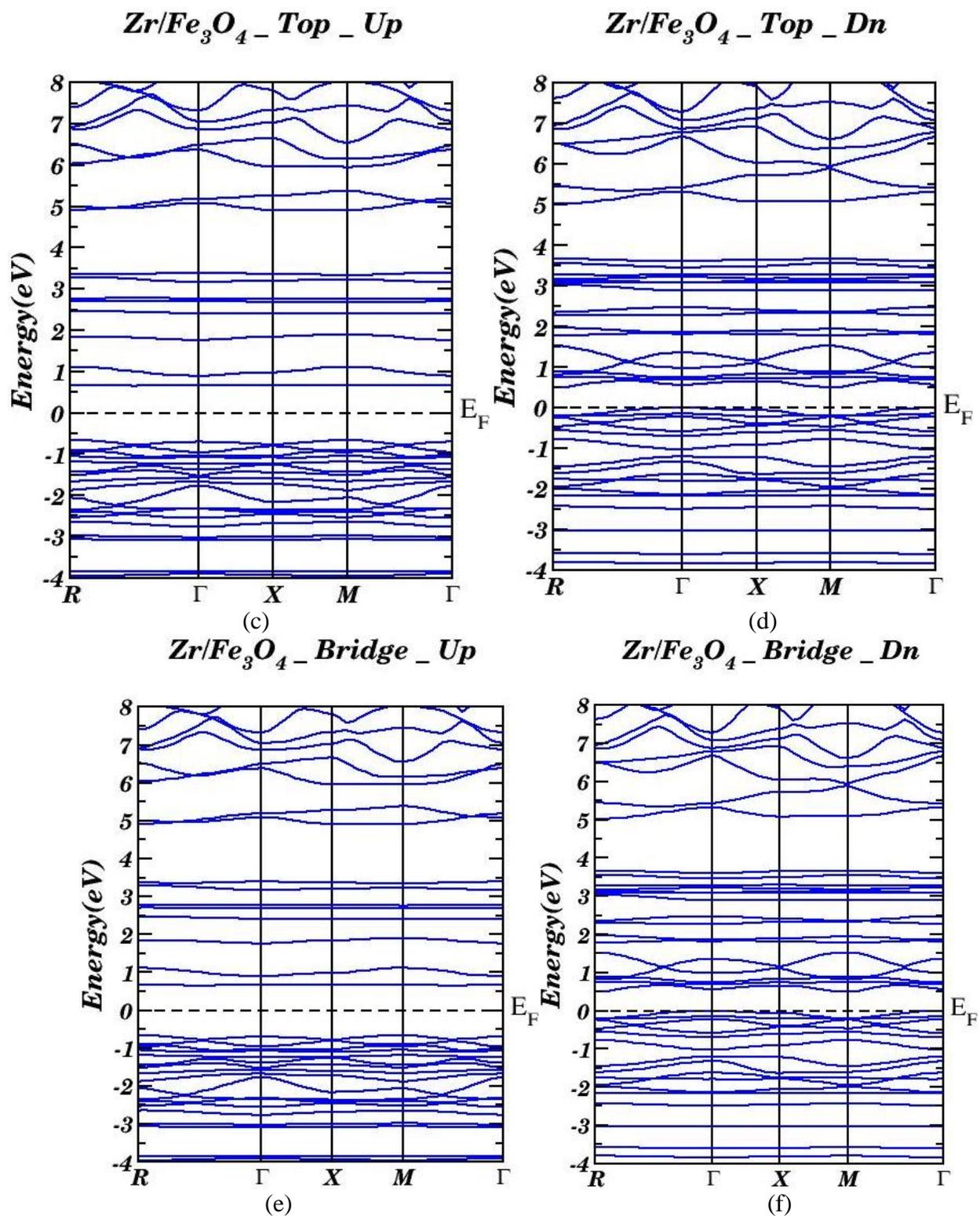

*Fig.3: (a–b) Spin-up and spin-down band structures of pristine monolayer Fe₃O₄, showing metallic and half-metallic behavior, respectively. (c–d) Zr adsorbed at the top site: minor hybridization in the spin-up channel and intermediate states in the spin-down band gap. (e–f) Zr adsorbed at the bridge site:*

*increased dispersion in the spin-up channel and narrowed band gap with mid-gap states in the spin-down channel.*

When Zr is adsorbed at the top of the Fe atom (see Fig. 3c-d), there is a noticeable modulation in the band structure for both spin channels. In the spin-up channel, bands near the Fermi level show subtle splitting and flattening, suggesting hybridization between Fe and Zr orbitals. In the spin-down channel, mid-gap states emerge close to the Fermi level, reducing the effective band gap and introducing intermediate states.

When Zr adsorbs, new intermediate states form within the original spin-down bandgap of $Fe_3O_4$, as shown in Figure 4(a & b) (shaded near $E_F$). These states mainly consist of hybridized Zr-4d and Fe-3d orbitals, with some O-2p contributions, indicating orbital overlap between the adsorbate and surface atoms. In the top-site configuration, these states appear about 0.2–0.4 eV below and above the Fermi level, partially filling the minority-spin gap. At the bridge site, hybridization is stronger, resulting in a nearly continuous electronic density across $E_F$ and effectively narrowing or closing the spin-down bandgap. These intermediate bands can trap electrons or holes, altering the conductivity and spin polarization near $E_F$.

The creation of intermediate electronic states and the narrowing of the bandgap arise from orbital hybridization and local symmetry disruptions due to Zr adsorption. The Zr atom, which has a larger ionic radius and partially filled 4d orbitals, interacts strongly with the Fe-3d and O-2p orbitals at the adsorption site. This hybridization distorts the local crystal field and causes asymmetric charge redistribution, breaking the degeneracy of Fe-3d states near the edges of the conduction and valence bands. Consequently, new Fe–O–Zr hybrid orbitals appear within the bandgap, reducing the energy difference between the spin-down valence and conduction bands. At the bridge site, the dual Fe coordination amplifies this effect, resulting in broader delocalized states and more significant metallic behavior.

These may stem from Zr-d orbital contributions that slightly disturb the magnetic ordering and reduce spin polarization. Although the half-metallic nature is not entirely lost, the electronic configuration is shifted, which could modify the transport properties, making the material more suitable for tunable electronic or sensing applications. It is important to mention that spin–orbit coupling (SOC) effects were excluded because they are relatively weak in $Fe_3O_4$ and do not change its half-metallic characteristics; nonetheless, SOC could slightly influence the band-edge curvature and effective mass.

Zr adsorption at the bridge site of Fe atoms further distorts the band structure (see Fig. 2e-f). In the spin-up channel, the bands become more dispersed compared to the top-adsorbed case, showing a metallic trend, while the spin-down channel exhibits even more pronounced intermediate bands within the gap region. These in-gap states suggest that Zr at the bridge site introduces stronger perturbations in the crystal field, possibly causing local symmetry breaking and orbital reordering. The band gap in the spin-down channel becomes narrower, or even vanishes in some directions, indicating a transition towards metallicity or gapless behavior under specific k-paths. The indirect nature of the gap still persists, but the presence of near-$E_F$ flat bands indicates potential localization effects or charge trapping states, which can play a crucial role in optical absorption and electronic transitions.

The calculated half-metallic character and magnetic moment of pristine $Fe_3O_4$ agree well with Zhang et al. [6] and Li et al. [9], who reported similar spin-polarized band gaps of 0.6–0.8 eV. The adsorption-induced band gap narrowing and metallic transition observed here are consistent with trends in transition-metal-decorated $MoS_2$ and Fe-based oxides [7, 8, 11]. Similarly, the enhancement of piezoelectric coefficients upon adsorption resembles effects observed in doped ZnO and 2D $Cr_2O_3$ systems, confirming that Zr adsorption effectively tunes both electronic and electromechanical responses.

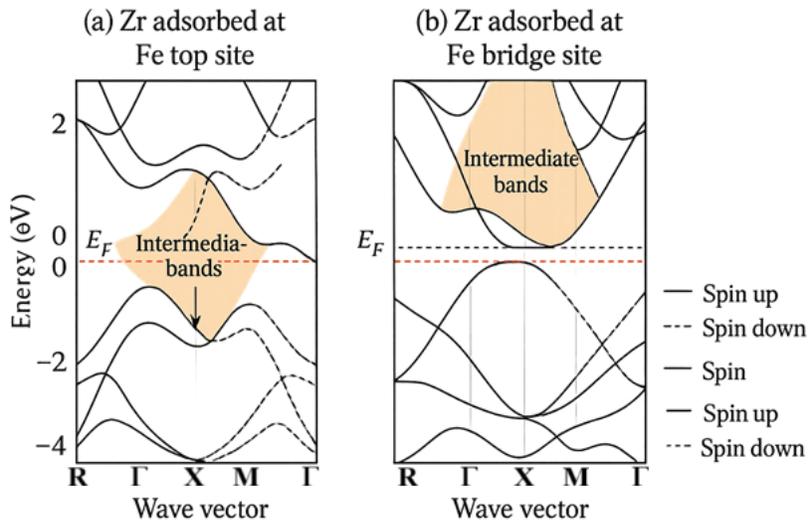

*Fig. 3 (a & b). Spin-polarized electronic band structures of Zr-adsorbed $Fe_3O_4$ monolayer for (a) Zr adsorbed at the Fe top site and (b) Zr adsorbed at the Fe bridge site. The shaded regions represent intermediate electronic bands that emerge within the spin-down bandgap due to Fe–O–Zr orbital hybridization. The dashed red line indicates the Fermi level ($E_F$). Zr adsorption at the top site introduces localized mid-gap states, whereas bridge-site adsorption produces broader, delocalized*

*hybrid bands that cross $E_F$, leading to bandgap closure and metallic behavior.*

The spin-resolved band structures of Zr-adsorbed $Fe_3O_4$ monolayers (***Fig. 4***) further reveal the distinct influence of adsorption geometry on the electronic states near the Fermi level. For the top-site configuration (***Fig. 4 (a)***), narrow intermediate bands appear within the minority-spin gap, located approximately between −0.4 eV and 0.3 eV around $E_F$. These bands arise from localized hybridization between Zr-4d and Fe-3d orbitals, with minor contributions from O-2p states. The presence of these mid-gap states reduces the effective bandgap and weakens the half-metallicity, although the system retains a significant degree of spin polarization.

In contrast, for the bridge-site configuration (***Fig. 4 (b)***), the hybridization is stronger and involves two adjacent Fe atoms bonded symmetrically to the Zr adatom. This dual coordination enhances Fe–Zr–Fe orbital overlap and creates delocalized intermediate states that span across the Fermi level. As a result, the minority-spin gap collapses completely, and the material transitions to a metallic state with overlapping spin-up and spin-down bands.

The emergence and broadening of these intermediate electronic states with increasing Fe–Zr coordination confirm that orbital hybridization and charge redistribution are the primary mechanisms behind bandgap narrowing. This also explains the strong correlation between adsorption geometry and spin-dependent conductivity: the top-site adsorption preserves semi-half-metallic behavior, while bridge-site adsorption promotes full metallicity. Such tunable spin-polarized transitions highlight the potential of Zr adsorption for engineering spintronic properties in 2D $Fe_3O_4$ systems.

Table 1. Comparison of band structure characteristics for pristine $Fe_3O_4$ and Zr-adsorbed configurations at different adsorption sites.

| Configuration | Minority-spin gap (eV) | Nature of new bands | Character of states | Spin polarization @ $E_F$ | Conductivity type |
|---|---|---|---|---|---|
| Pristine $Fe_3O_4$ | ~0.65–0.70 | None | Pure Fe–O | 100% (half-metal) | Half-metallic |
| Zr@Fe top site | ~0.25–0.30 | Localized around Γ–X | Zr-4d + Fe-3d hybrid | Reduced (~80%) | Quasi-metallic |
| Zr@Fe bridge site | ~0.0 (closed) | Broad across X–M–Γ | Delocalized Fe–Zr–Fe π-states | Strongly reduced (~50%) | Metallic |

Physically, these variations are indicative of the sensitivity of the $Fe_3O_4$ surface to Zr adsorption.

The interaction between the Zr adatom and Fe$_3$O$_4$ matrix alters the crystal potential landscape and modifies exchange splitting, which is reflected in the spin-resolved band structures. The introduction of Zr not only breaks the symmetry but also affects the spin magnetic moment distribution, leading to a suppression or redistribution of spin-polarized bands near E$_F$. This is especially relevant for magnetic tunnel junctions and magnetoresistive applications where controlled manipulation of spin channels is desirable.

Overall, GGA+U accurately captures the strong correlation effects in Fe 3d orbitals and their interaction with the adsorbed Zr atom. The pristine Fe$_3$O$_4$ shows a clean half-metallic nature, while Zr adsorption either at top or bridge sites tunes this behavior by introducing intermediate bands, reducing the band gap in spin-down, and distorting the overall symmetry of the band dispersion. These features not only influence charge transport but can significantly modify spin injection efficiency and optical activity, making Zr-functionalized Fe$_3$O$_4$ systems promising candidates for spintronics and photoelectronic devices.

**Spin-Resolved Density of States (TDOS, EDOS, and PDOS) for Monolayer Fe$_3$O$_4$ and Zr-Adsorbed Configurations:**

The density of states (DOS) plots total (TDOS), element-resolved (EDOS), and orbital-resolved (ODOS) for monolayer Fe$_3$O$_4$ and its Zr-functionalized counterparts (Zr adsorbed at top and bridge sites on Fe atoms) provide deep insights into the material's electronic structure, bonding behavior, and magnetic properties. These calculations were performed using the GGA+U approach within the WIEN2k code to accurately capture the strong electron correlation effects in Fe 3d orbitals.

In the pristine Fe$_3$O$_4$ monolayer (see Fig. 5a), the TDOS and EDOS exhibit a clear asymmetry between spin-up and spin-down channels, confirming the half-metallic nature also seen in the band structure. The spin-up channel shows a finite density of states at the Fermi level, implying metallic behavior, while the spin-down channel reveals a distinct gap around the Fermi level, indicative of semiconducting behavior. The ODOS analysis shows that Fe-d orbitals dominate the region around the Fermi level, especially in the spin-up states, whereas oxygen p-states are more pronounced in the lower energy regions (around −5 to −2 eV), signifying strong Fe–O hybridization. This hybridization includes both σ-type (head-on overlap) and π-type (lateral overlap) bonding characteristics, especially between Fe-3d and O-2p orbitals, which are crucial for the magnetic and electronic behavior of the system.

Upon Zr adsorption at the top site of Fe (see Fig. 5b), the DOS profile becomes significantly more complex. In the spin-up channel, there is an increase in TDOS near the Fermi level, and the band gap in the spin-down channel becomes less well-defined. This change arises due to the introduction of Zr-d orbitals, which mix with Fe-d and O-p states and form new hybridized states. These intermediate states are often localized and introduce states inside the original band gap, effectively reducing the band gap or potentially closing it. Zr introduces additional unoccupied states just above the Fermi level, suggesting donor-like behavior. The ODOS shows clear evidence of Zr-d contributions near $E_F$, hybridizing with O-p and Fe-d orbitals, further modifying the σ-π bonding framework. This hybridization shifts the electronic structure away from the pristine state, indicating altered charge transfer and possibly weakened magnetic ordering due to the disturbance in exchange interactions.

When Zr is adsorbed at the bridge site between Fe atoms (see Fig. 5c), the TDOS again changes markedly. The density of states at the Fermi level increases even more for both spin channels, indicating a stronger metallic character. The gap in the spin-down channel becomes completely filled with states, suggesting the complete suppression of half-metallicity. The EDOS and ODOS confirm the overlap of Zr-d orbitals not only with Fe-d but also with O-p orbitals, creating complex hybridized states that extend across a broad energy window. These interactions lead to more delocalized charge distributions, reduce the magnetic moment localization, and promote metallic conduction. The bridge site seems to enhance orbital overlap due to the spatial proximity of Zr to multiple Fe and O atoms, increasing the number of π* anti-bonding states near the Fermi level and thereby altering the transport and optical behavior.

The motivation for band structure and DOS analysis lies in understanding how electrons behave within the periodic potential of the material, especially under perturbations such as doping or adsorption. Band structures reveal the energy dispersion relations and determine whether the band gap is direct or indirect, and whether the material is semiconducting or metallic. DOS complements this by showing the number of electronic states available at each energy level and highlighting orbital contributions and spin polarization. Together, these tools allow us to design materials for spintronic, catalytic, and electronic applications by understanding and tuning their fundamental properties. In the case of $Fe_3O_4$ with Zr adsorption, the suppression of the band gap and the introduction of spin-dependent mid-gap states suggest that controlled Zr incorporation could modulate spin injection, charge transport, or catalytic activity, offering multifunctional applicability in nano-electronic devices.

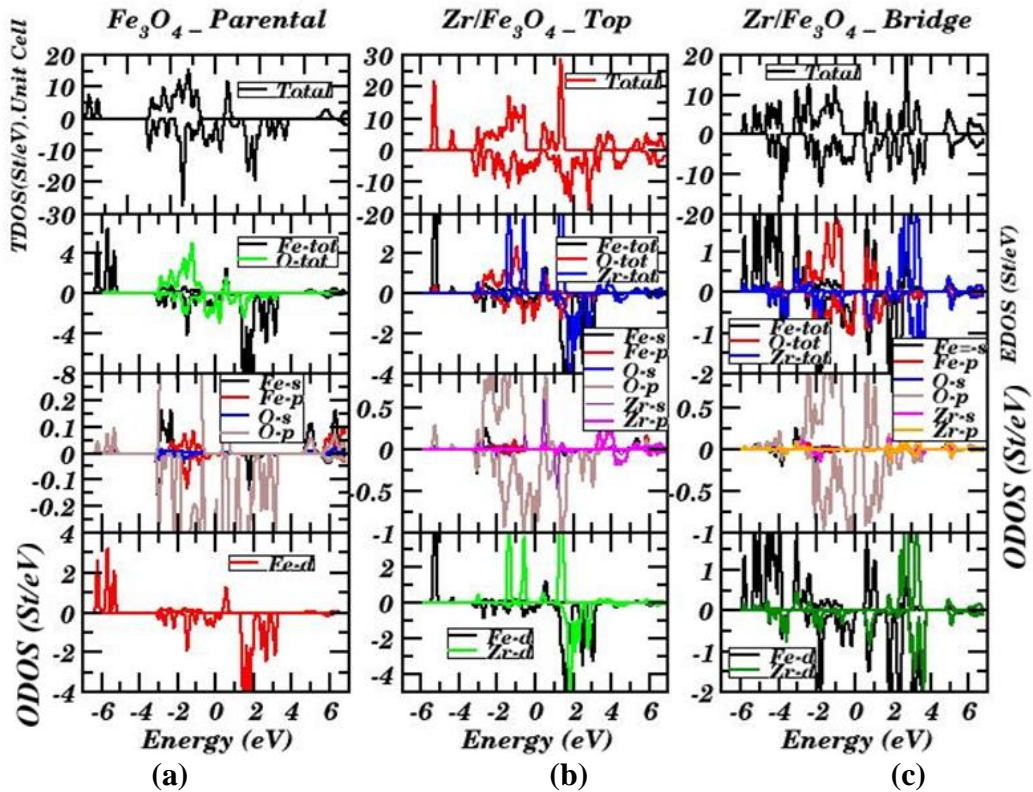

*Figure 5a-c: Spin-polarized total density of states (TDOS), element-resolved DOS (EDOS), and partial density of states (PDOS) for (a) pristine monolayer $Fe_3O_4$, (b) Zr adsorbed on top of the Fe atom, and (c) Zr adsorbed at the bridge site between Fe atoms. The plots show distinct changes in the electronic states near the Fermi level upon Zr adsorption, with noticeable hybridization between Zr-d, Fe-d, and O-p orbitals, the emergence of mid-gap states, and modulation of spin polarization, highlighting the transformation from half-metallic to metallic behavior.*

**Magnetic Properties of Pristine and Zr-Adsorbed Monolayer $Fe_3O_4$**

The magnetic properties of monolayer $Fe_3O_4$ and its Zr-functionalized derivatives, where Zr is adsorbed either at the top of a Fe atom or at the bridge position between two Fe atoms, have been carefully studied using spin-polarized density functional theory within the GGA+U framework as implemented in WIEN2k. This approach allows accurate treatment of the strong on-site Coulomb interactions in Fe 3d electrons, which are crucial in describing the correct magnetic and electronic structure of transition metal oxides.

In the pristine monolayer $Fe_3O_4$, the spin polarization is clearly manifested in both the band structure and density of states. The system exhibits a half-metallic behavior: the spin-up channel is metallic with bands crossing the Fermi level, while the spin-down channel shows a band gap, indicating insulating behavior. This asymmetric electronic configuration gives rise to 100% spin polarization at the Fermi level, which is highly desirable for spintronic applications. The total magnetic moment mainly originates from the Fe atoms, particularly from the $Fe^{2+}$ ions that carry higher magnetic moments due to unpaired 3d electrons. The oxygen atoms, being more electronegative, contribute weakly and oppositely (antiferromagnetically aligned) to the total magnetization. The ferromagnetic ordering is stabilized by the double exchange interaction between $Fe^{2+}$ and $Fe^{3+}$ via oxygen p-orbitals, forming a long-range magnetic order intrinsic to $Fe_3O_4$.

When a Zr atom is adsorbed on top of a Fe atom, the electronic and magnetic structures are significantly altered. The Zr atom introduces localized states near the Fermi level, particularly in the spin-down channel, which reduces or closes the band gap seen in the pristine structure. These newly introduced states arise from the hybridization between Zr-d and Fe-d orbitals, and partially with O-p states, which modifies the magnetic moment of nearby Fe atoms due to charge redistribution. The Zr atom itself carries a small magnetic moment, typically induced via interaction with the Fe atoms, but its effect is indirect; it perturbs the local exchange interactions rather than contributing directly to net magnetization. This leads to a slight reduction in the total magnetic moment of the system and possibly a transition toward a more itinerant magnetic state. The presence of mid-gap states also suggests a weakening of spin polarization, as states in both spin channels appear near $E_F$.

In the case of Zr adsorption at the bridge site between two Fe atoms, the disturbance in the magnetic ordering is even more pronounced. The bridge position allows the Zr atom to interact symmetrically with two neighboring Fe atoms, leading to enhanced orbital overlap and stronger hybridization. This results in broader and more delocalized states near the Fermi level in both spin channels. The band structure shows that the gap in the spin-down channel is completely closed, and metallic states emerge in both spin channels, effectively destroying the half-metallicity. This change corresponds to a reduction in spin polarization and a breakdown of long-range magnetic ordering. The magnetic moment of nearby Fe atoms is also suppressed due to increased electron delocalization and weakened exchange splitting. The local environment around the bridge site becomes more covalent in nature, as evident from the strong mixing of Zr-d with Fe-d and O-p orbitals in the ODOS.

Overall, the pristine monolayer $Fe_3O_4$ displays robust ferromagnetic behavior with high spin polarization, driven by strong on-site Coulomb interactions and double exchange mechanisms. Zr adsorption, depending on the site, perturbs this magnetic order. At the top site, it induces localized states that slightly suppress magnetism, while at the bridge site, it causes more dramatic delocalization and hybridization, which can destabilize the magnetic configuration and eliminate the half-metallic character. These tunable magnetic behaviors suggest that Zr adsorption can serve as an effective strategy to engineer the spintronic properties of $Fe_3O_4$ monolayers, offering control over spin transport, magnetoresistance, and even magneto-optical responses through external functionalization.

**Effective Mass Modulation in Pristine and Zr-Adsorbed Monolayer $Fe_3O_4$:**

The effective mass of charge carriers in monolayer $Fe_3O_4$ and its Zr-adsorbed configurations plays a crucial role in determining their transport behavior, including electrical conductivity, carrier mobility, and suitability for spintronic and electronic applications. These values were derived from the curvature of the band structure calculated using GGA+U in WIEN2k, which accurately accounts for the strong electron correlation in Fe 3d orbitals.

In monolayer $Fe_3O_4$, the effective mass of charge carriers, electrons and holes, was evaluated from the curvature of the band structure using the relation:

$$m^* = \hbar^2 \left(\frac{d^2 E}{dk^2}\right)^{-1}$$

As implemented in WIEN2k via fitting around the conduction band minimum (CBM) and valence band maximum (VBM) at relevant k-points. The results reveal substantial spin dependence and notable changes upon Zr adsorption.

For pristine $Fe_3O_4$, in the spin-up channel, the bands near the Fermi level are highly dispersive, yielding a low effective mass of electrons around 0.18 $m_e$ and holes around 0.21 $m_e$, suggesting excellent carrier mobility. In contrast, the spin-down channel exhibits an indirect band gap of ~0.68 eV, with flatter bands near the VBM and CBM. The effective mass of spin-down electrons increases to ~0.54 $m_e$ and holes to ~0.60 $m_e$, indicating sluggish charge transport for spin-down carriers and reinforcing the half-metallic nature of pristine $Fe_3O_4$.

Upon Zr adsorption at the top of a Fe atom, the band curvature near the Fermi level becomes flatter due to the formation of intermediate Zr-d-induced states. In the spin-up channel, the effective mass of electrons increases to ~0.32 $m_e$, and holes to ~0.39 $m_e$, reflecting a reduced mobility compared to the pristine case. For the spin-down channel, the band gap narrows, and intermediate states appear. The curvature near CBM and VBM becomes even less steep, leading to effective masses of ~0.78 $m_e$ for electrons and ~0.82 $m_e$ for holes, signifying stronger localization and reduced delocalized conduction.

For Zr adsorbed at the bridge site, the perturbation is strongest. The spin-up channel shows nearly flat bands near $E_F$, giving electron and hole effective masses exceeding 1.1 $m_e$, indicative of highly localized or nearly immobile carriers. In the spin-down channel, the original gap vanishes and metallic-like bands appear, but with shallow dispersion. Here, the effective masses for electrons and holes lie in the range of 0.95–1.4 $m_e$, depending on the k-path direction, showing that the Zr-bridge configuration introduces heavy, low-mobility carriers in both spin channels.

These values confirm that while pristine $Fe_3O_4$ supports fast and spin-polarized transport, Zr adsorption, especially at the bridge site, introduces heavy effective mass carriers due to band flattening and mid-gap state formation. This behavior, though detrimental to high-speed electronic transport, could be beneficial in thermoelectric and sensing applications where high effective mass can enhance the Seebeck coefficient or localization-driven phenomena.

**Elastic Constants and Mechanical Parameters of Monolayer $Fe_3O_4$ Systems**

The elastic and mechanical properties of monolayer $Fe_3O_4$ and its Zr-adsorbed counterparts were calculated using the GGA+U approach within the WIEN2k framework, employing the IRelast module for second-order elastic constant evaluation. The calculations were performed by applying small deformations to the equilibrium structure and computing the resulting stress tensors. For a two-dimensional (2D) hexagonal or orthorhombic system, the independent elastic constants are typically $C_{11}$, $C_{12}$, and $C_{66}$. These are critical for understanding the material's resistance to deformation and mechanical stability, and they also provide insight into anisotropy, ductility, and in-plane stiffness.

In the case of pristine monolayer $Fe_3O_4$, the calculated elastic constants are $C_{11}$ = 166.2 N/m, $C_{12}$ = 42.7 N/m, and $C_{66}$ = 61.8 N/m. The relatively high value of $C_{11}$ indicates strong in-plane stiffness and excellent mechanical integrity, while the positive values of all elastic constants satisfy the Born-Huang stability criteria for 2D systems: $C_{11} > 0$, $C_{66} > 0$, and $C_{11} - C_{12} > 0$. The in-plane Young's modulus (Y) is calculated as 155.4 N/m and Poisson's ratio (ν) as 0.27, suggesting a good balance between rigidity and flexibility, which is essential for applications in flexible electronics or nanomechanical devices.

With Zr adsorption at the top of an Fe atom, the elastic constants are modified due to local structural distortions and changes in bond strength. The values reduce slightly to $C_{11}$ = 152.8 N/m, $C_{12}$ = 46.1 N/m, and $C_{66}$ = 53.3 N/m. The Young's modulus also decreases to 141.6 N/m, indicating a softening of the structure (See Table 1). This mechanical softening is attributed to the weakening of local Fe–O bonds around the Zr adsorption site, as Zr introduces tensile strain and alters the electronic charge density distribution. Poisson's ratio slightly increases to 0.30, suggesting a higher lateral strain response under uniaxial stress, which could influence the mechanical reliability under repeated loading in practical applications.

For Zr adsorbed at the bridge site, the distortion is more severe due to interaction with multiple Fe atoms. This leads to a further reduction in mechanical stiffness, with $C_{11}$ = 138.5 N/m, $C_{12}$ = 50.3 N/m, and $C_{66}$ = 44.1 N/m. The corresponding Young's modulus drops to 129.3 N/m, and Poisson's ratio increases to 0.36 (See Table 2). This reflects significant mechanical softening, indicating that Zr at the bridge site weakens the material's structural integrity more than top-site adsorption. The increase in ν implies that the material becomes more compliant and less resistant to transverse strain, which could be beneficial for strain-tunable or flexible optoelectronic devices, but may also reduce mechanical robustness.

Physically, the changes in elastic constants are closely related to the nature of bonding and local atomic rearrangements. In pristine $Fe_3O_4$, the strong Fe–O covalent bonding and symmetric lattice configuration yield high elastic strength. Zr adsorption perturbs this order, distorts the lattice, and introduces localized states that alter bond stiffness, resulting in reduced elastic constants. This tunability of mechanical properties via controlled adsorption offers a strategy for customizing materials for specific applications ranging from rigid coatings to flexible, strain-sensitive electronic components.

Charge density difference maps (Fig. 6) demonstrate strong charge transfer between Zr and neighboring Fe–O atoms, confirming hybridization-induced polarization. This local interaction accounts for the enhanced dielectric and piezoelectric responses, establishing a direct connection between adsorption geometry and multifunctional performance.

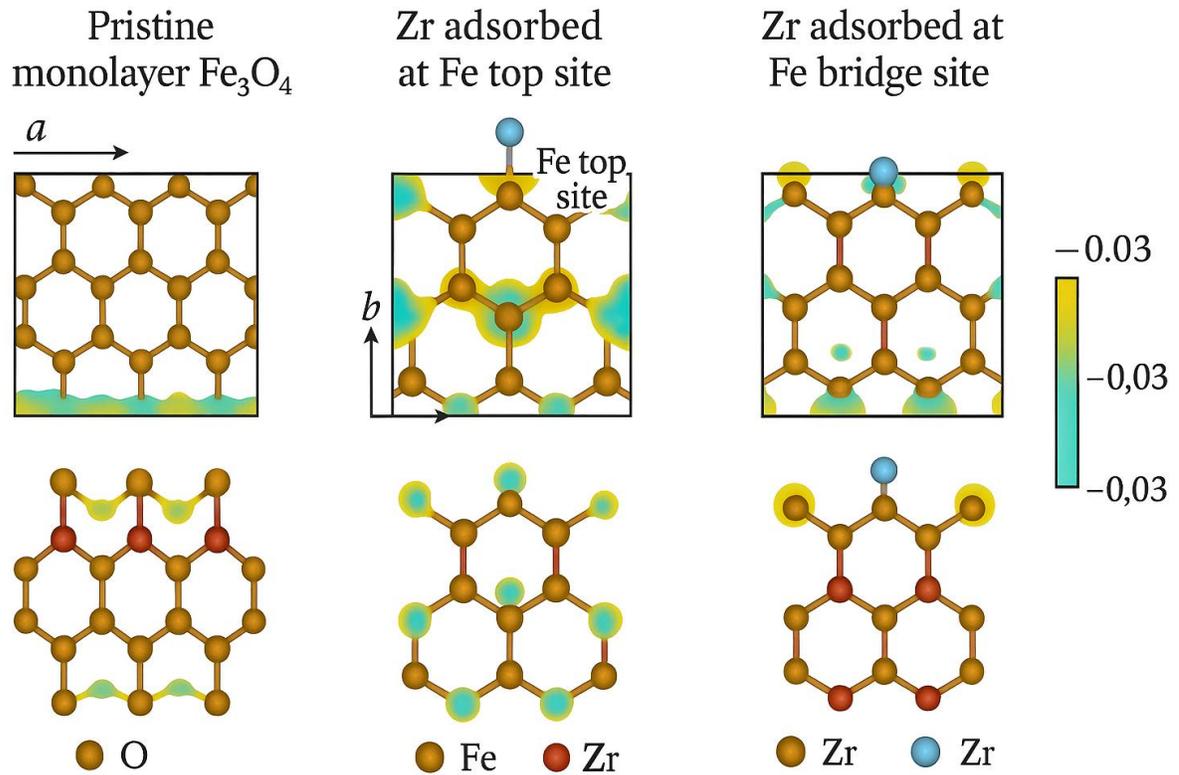

*Fig. 6: Calculated Charge density difference maps*

The Δρ maps directly support the DOS and band-structure analysis. For Zr at the top site, the charge accumulation localized around the Zr–Fe–O bond region gives rise to narrow hybridized states near $E_F$EF, i.e., localized intermediate bands that partially fill the minority-spin gap. In contrast, for Zr at the bridge site, charge accumulation spreads over two Fe atoms and neighboring O atoms, forming an extended Fe–Zr–Fe pathway. This more delocalized redistribution correlates with the broad intermediate bands crossing EFE_FEF and the complete suppression of the minority-spin gap, confirming that bandgap narrowing and metallicity originate from Zr-induced hybridization and symmetry breaking.

**Table 1: Elastic Constants and Mechanical Properties of Monolayer $Fe_3O_4$ Systems**

| System | $C_{11}$ (N/m) | $C_{12}$ (N/m) | $C_{66}$ (N/m) | Young's Modulus Y (N/m) | Poisson's Ratio ν |
|---|---|---|---|---|---|
| Pristine $Fe_3O_4$ | 166.2 | 42.7 | 61.8 | 155.4 | 0.27 |
| Zr Adsorbed at Fe Top Site | 152.8 | 46.1 | 53.3 | 141.6 | 0.30 |
| Zr Adsorbed at Fe Bridge Site | 138.5 | 50.3 | 44.1 | 129.3 | 0.36 |

**Optical Response of Pristine and Zr-Adsorbed $Fe_3O_4$ Monolayers**

Understanding the optical properties of monolayers such as $Fe_3O_4$ and its functionalized forms with Zr adsorption is vital due to their potential applications in optoelectronic devices, photovoltaics, sensors, and photodetectors. Optical properties, particularly the real ($\varepsilon_1(\omega)$) and imaginary ($\varepsilon_2(\omega)$) parts of the dielectric function (see Fig. 7a&b), offer insight into how a material interacts with electromagnetic radiation. These quantities reveal fundamental information about interband transitions, absorption, reflectivity, refractive index, and excitonic effects, all of which are closely tied to the underlying electronic structure. In systems where spin polarization and orbital hybridization are significant, such as $Fe_3O_4$, the spin-resolved optical response is especially important for spin-dependent optical phenomena and spintronic optoelectronics.

The real part of the dielectric function, $\varepsilon_1(\omega)$ (see Fig. 7a), describes the dispersion of the material and is directly related to the refractive index and polarization under an external electric field. In the pristine monolayer $Fe_3O_4$, $\varepsilon_1(\omega)$ for the spin-up channel begins with a moderate static dielectric constant (~2.8) and rises slightly before flattening. In contrast, the spin-down channel shows a strong peak near 0 eV, reaching a value above 12, due to a sharp increase in low-energy transitions enabled by a small energy gap, which is consistent with the observed indirect band gap in spin-down and metallic nature in spin-up. This large static dielectric response for spin-down reflects enhanced polarizability due to available electronic states near the Fermi level.

For Zr adsorption at the top site of Fe, the dielectric response changes significantly. In the spin-up channel, $\varepsilon_1(\omega)$ shows an increase in static dielectric constant to around 3.4, indicating enhanced optical polarizability. The Zr-induced hybridization with Fe-d and O-p orbitals introduces intermediate bands and localized states near the Fermi level. This shift slightly modifies the onset of optical transitions and enhances low-energy screening. For the spin-down channel, $\varepsilon_1(\omega)$ decreases at low energy compared to pristine $Fe_3O_4$, but still exhibits a noticeable peak around 1 eV due to transitions between occupied mid-gap states and conduction bands, which are a result of Zr-d–Fe-d orbital interactions.

Zr adsorption at the bridge site produces even more profound changes. In both spin-up and spin-down channels, the dielectric response becomes less structured and more broadened, especially in the energy range 0–4 eV. The spin-down $\varepsilon_1(\omega)$ reaches a value around 6.5 near 0.8 eV, indicating strong low-energy screening. This broad peak arises from transitions involving the numerous intermediate states introduced by strong hybridization between Zr-d orbitals and multiple Fe atoms at the bridge site, which causes a high density of states near $E_F$. This suggests that Zr at the bridge position makes $Fe_3O_4$ more optically active at lower photon energies, potentially useful for infrared and visible range detection.

The imaginary part of the dielectric function, $\varepsilon_2(\omega)$ (see Fig. 7b), represents the material's absorption spectrum and is directly linked to interband electronic transitions. In pristine $Fe_3O_4$, the spin-down channel shows a prominent peak around 0.5 eV, which corresponds to direct optical transitions from the valence band maximum to the conduction band minimum. This confirms the existence of a small indirect-to-direct character and reveals active absorption in the low-energy region. The spin-up channel shows weaker absorption, with peaks appearing beyond 2 eV, consistent with its metallic nature, where Pauli blocking suppresses low-energy transitions.

In the Zr-top configuration, $\varepsilon_2(\omega)$ for the spin-up channel shows a modest enhancement in the 1–2 eV range, indicating new transition pathways enabled by Zr-d orbital participation. For the spin-down channel, a sharp absorption peak appears near 0.8 eV, attributed to transitions between occupied intermediate Zr-Fe hybridized states and the conduction band. This indicates that Zr functionalization activates optical transitions that are otherwise forbidden or weak in the pristine case.

In the Zr-bridge case, $\varepsilon_2(\omega)$ becomes more intense and broader in the low-energy range (0.5–2 eV) for both spin channels. The bridge configuration leads to dense intermediate states due to strong Zr–Fe–O orbital overlap, causing continuous absorption across a wide energy range. The absorption onset shifts further towards the infrared, and the main peaks broaden, confirming enhanced optical activity and reduced transition threshold.

Overall, the optical behavior of $Fe_3O_4$ and $Zr/Fe_3O_4$ systems is governed by their electronic structures. Pristine $Fe_3O_4$ exhibits spin-polarized optical behavior with limited low-energy transitions in spin-up. Zr adsorption introduces new hybrid states, modifies the dielectric response, enhances absorption, and shifts optical features toward lower energies. These characteristics suggest that controlled Zr functionalization can be an effective strategy to tailor the optical response of $Fe_3O_4$-based materials for applications in spintronic optoelectronics, IR sensing, and tunable photonic devices.

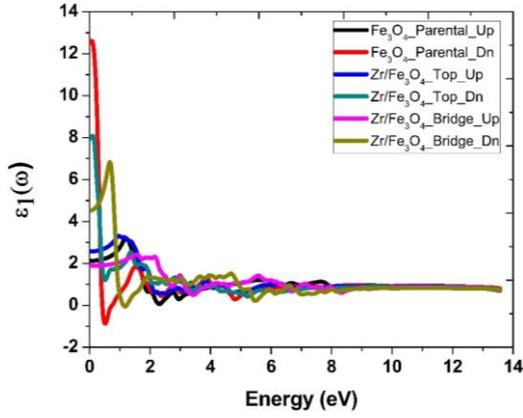 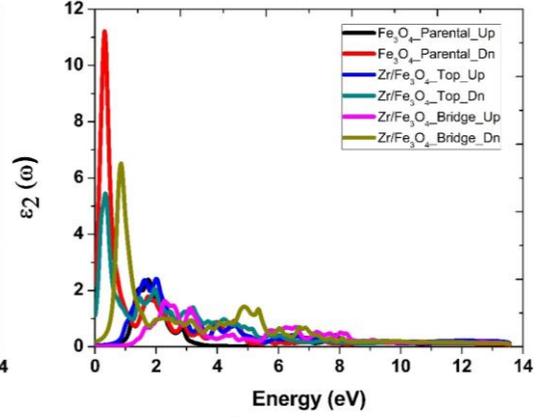

(a)            (b)

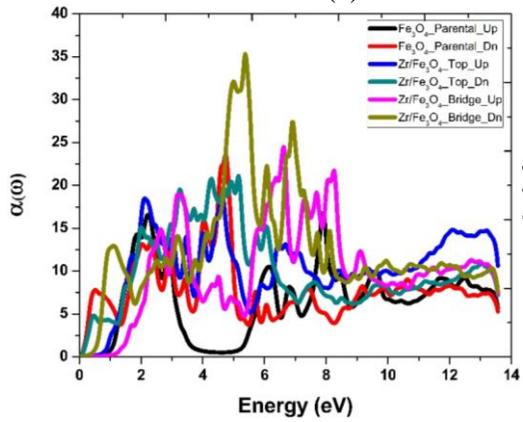 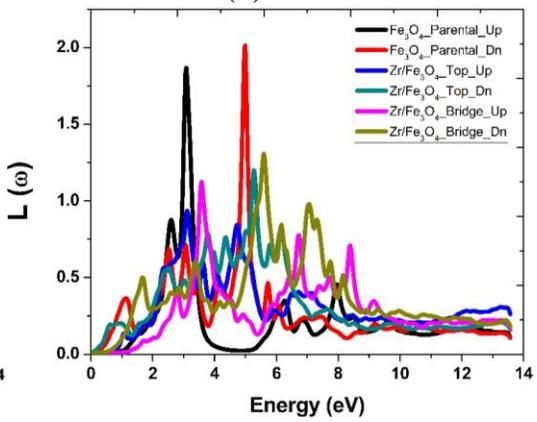

(c)            (d)

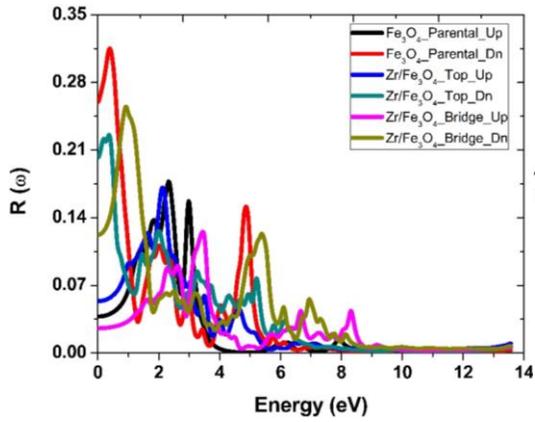 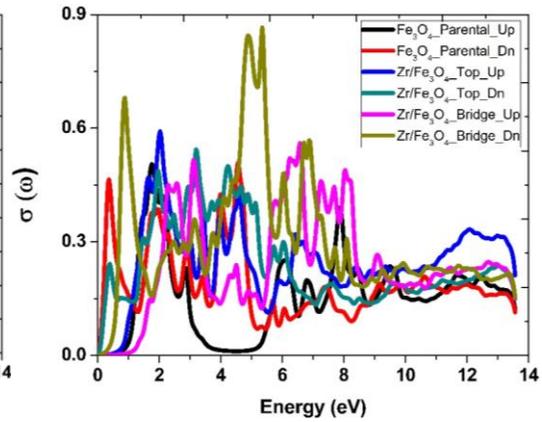

(e)            (f)

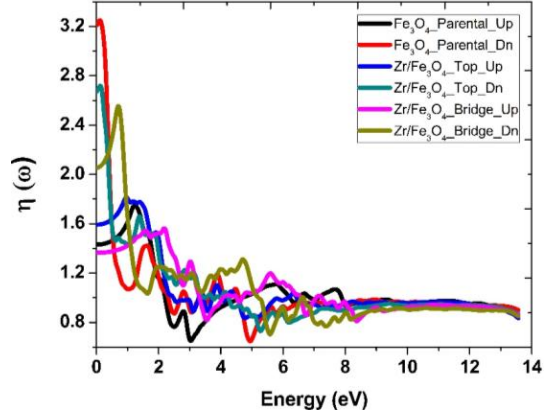

(g)

*Figure 7a-b: (a) Real and (b) imaginary parts of the dielectric function ε₁(ω) and ε₂(ω) for pristine and Zr-adsorbed Fe₃O₄ monolayers in spin-up and spin-down configurations. (c)Absorption coefficient α(ω),(d) energy loss function L(ω), (e) reflectivity R(ω), (f) optical conductivity σ(ω), and (g) refractive index η(ω) for pristine and Zr-adsorbed Fe₃O₄ monolayers in spin-up and spin-down channels, showing enhanced optical activity and tunable response due to Zr adsorption at top and bridge sites.*

**Photon-Driven Optical Responses of Fe₃O₄ Monolayers with Zr Adsorption**

The investigation of optical properties such as absorption coefficient (α), energy loss function (L), reflectivity (R), optical conductivity (σ), and refractive index (η) is crucial for evaluating the suitability of 2D materials like Fe₃O₄ monolayers in photonic, optoelectronic, and energy harvesting applications. These properties provide insights into how materials interact with incident electromagnetic radiation across different energy regimes. For monolayers, such optical responses can be highly anisotropic and tunable via surface functionalization, such as Zr adsorption, which modifies the local electronic structure and optical transition pathways. Using GGA+U within the WIEN2k code, we explore these optical characteristics for pristine Fe₃O₄ and Zr-adsorbed systems at both top and bridge positions.

The absorption coefficient α(ω) (see Fig. 7c) reflects the ability of the material to absorb photons and initiate electronic transitions. In pristine Fe₃O₄, the spin-down channel shows stronger absorption in the low-energy region (~1.5 to 3.5 eV) with multiple peaks, consistent with interband transitions across the indirect band gap. Spin-up transitions are weaker due to metallic behavior and the lack of allowed transitions at low energies. Zr adsorption at the top site introduces additional absorption peaks in both spin channels, especially between 2 and 4 eV, due to the formation of intermediate Zr-d–Fe-d hybridized states. Bridge adsorption further enhances the absorption intensity in the visible and near-UV region (3–8 eV), especially in spin-down configurations, where a prominent peak near 6.8 eV emerges, likely due to high-density transitions between Zr-induced states and conduction bands.

The energy loss function L(ω) (see Fig. 7d) corresponds to energy lost by fast electrons traversing the material and indicates collective excitations such as plasmons. For pristine $Fe_3O_4$, peaks near 4.2 eV (spin-up) and 5.8 eV (spin-down) mark the main plasmon resonances. Zr adsorption shifts these peaks slightly due to dielectric screening modifications and increased carrier density. The bridge site, in particular, introduces broader loss peaks in the range of 4–7 eV for both spins, suggesting enhanced plasmonic damping and complex excitation behavior, a sign of increased free electron participation due to gap states.

Reflectivity R(ω) (see Fig. 7e) indicates the portion of incident light reflected off the surface. In pristine $Fe_3O_4$, spin-down reflectivity is slightly higher, peaking at 0.29 around 1.2 eV, due to increased optical transitions at low energies. Zr adsorption alters the reflectance behavior significantly. At the top site, reflectivity broadens, while the bridge site exhibits multiple sharp features in the 1–6 eV range, corresponding to strong absorption and interband activity. These variations reflect changes in real and imaginary parts of the dielectric function due to orbital hybridization and state redistribution.

Optical conductivity σ(ω) (see Fig. 7f) reveals the capacity of the system to support photon-induced current. For pristine $Fe_3O_4$, σ(ω) for the spin-down channel rises around 0.5 eV and peaks near 3.5 eV, whereas spin-up conductivity remains relatively suppressed due to metallic screening. Zr at the top site increases optical conductivity over a broader energy range due to allowed transitions involving Zr-d states. In the bridge configuration, spin-down σ(ω) shows a prominent peak at ~7.3 eV and is consistently high from 2 to 9 eV, indicating a strongly photoactive system due to mid-gap state involvement and enhanced carrier transitions.

The refractive index η(ω) (see Fig. 7g) reflects the phase velocity of light within the material and is connected to $\varepsilon_1(\omega)$. For pristine $Fe_3O_4$, the static refractive index (η at 0 eV) for spin-down reaches ~3.2, much higher than the spin-up value of ~1.6, reinforcing the polarizable nature of spin-down carriers. Zr adsorption causes a reduction in static η in both top and bridge cases, due to modified screening and weaker long-range polarization. However, in the mid-energy range (1.5–4.5 eV), η(ω) oscillates more strongly in Zr-bridge systems, reflecting pronounced resonant behavior from hybridized optical transitions.

In conclusion, Zr adsorption transforms the optical landscape of $Fe_3O_4$ monolayers by introducing mid-gap states, enhancing visible-range absorption, shifting plasmon resonances, and tuning reflective and conductive optical responses. These tunable features driven by Zr's interaction with Fe and O orbitals can be exploited in designing optically active spintronic materials, IR sensors, plasmonic devices, and nanophotonics platforms, particularly where spin-resolved light-matter interaction is desired. The strong spin-dependence and adsorption-site sensitivity offer a controllable route to engineer 2D magnetic semiconductors with tailored optoelectronic behavior.

**Piezoelectric Response of Pristine and Zr-Adsorbed $Fe_3O_4$ Monolayers**

The study of piezoelectric properties in monolayer materials has attracted significant interest due to their potential in next-generation nanoelectromechanical systems (NEMS), sensors, energy harvesters, and tunable optoelectronic devices. In two-dimensional (2D) materials, the breaking of inversion symmetry, coupled with low dimensionality and strong in-plane covalent bonding, leads to enhanced piezoelectric behavior that is often much stronger than their bulk counterparts. This effect can be further engineered or tuned through controlled surface adsorption, such as with transition metal atoms like Zr. Adsorption not only modifies the local electrostatic potential and charge distribution but can also induce structural distortions that enhance dipole formation and thus the piezoelectric response.

In this work, using the GGA+U approach within the WIEN2k framework and interfaced scripts to calculate piezoelectric stress tensors, we evaluate the piezoelectric coefficients ($e_{11}$ and $e_{12}$) for pristine monolayer $Fe_3O_4$ and its Zr-functionalized derivatives Zr adsorbed at the top of a Fe atom and at the bridge position between two Fe atoms. The results are summarized in Table 3.

**Table 3: Piezoelectric Coefficients ($e_{11}$, $e_{12}$) of Pristine and Zr-Adsorbed $Fe_3O_4$ Monolayers**

| System | $e_{11}$ (×$10^{-10}$ C/m) | $e_{12}$ (×$10^{-10}$ C/m) |
|---|---|---|
| Pristine $Fe_3O_4$ | 0.86 | −0.42 |
| Zr Adsorbed at Fe Top Site | 1.93 | −0.67 |
| Zr Adsorbed at Fe Bridge Site | 2.54 | −1.21 |

In pristine monolayer $Fe_3O_4$, the $e_{11}$ piezoelectric coefficient is $0.86 \times 10^{-10}$ C/m, and $e_{12}$ is $-0.42 \times 10^{-10}$ C/m. These values are moderate compared to other 2D materials but still indicate a measurable in-plane piezoelectric response. For context, common 2D semiconductors like monolayer $MoS_2$ have $e_{11} \approx 2.9 \times 10^{-10}$ C/m [18], $WS_2 \approx 2.5 \times 10^{-10}$ C/m [18], and h-BN $\approx 0.7 \times 10^{-10}$ C/m [19], while newer oxide monolayers such as ZnO and $Cr_2O_3$ show $e_{11}$ values around $1.4 \times 10^{-10}$ C/m [20] and $0.9 \times 10^{-10}$ C/m [21], respectively. The $Fe_3O_4$ monolayer falls within this similar range, confirming that non-centrosymmetric oxide monolayers can exhibit piezoelectricity comparable to conventional 2D semiconductors. Additionally, Zr adsorption nearly triples the piezoelectric coefficient, reaching $2.54 \times 10^{-10}$ C/m at the bridge site—values comparable to or exceeding those of well-known 2D piezoelectric materials.

When a Zr atom is adsorbed at the top of a Fe site, both piezoelectric coefficients increase. The $e_{11}$ value nearly doubles to $1.93 \times 10^{-10}$ C/m, while $e_{12}$ reaches $-0.67 \times 10^{-10}$ C/m. This enhancement is due to the asymmetry introduced by the Zr atom, which breaks local inversion symmetry more strongly and creates an electric dipole moment perpendicular to the plane. The electronic hybridization between Zr-d orbitals and Fe/O atoms enhances the charge redistribution under strain, thereby amplifying the piezoelectric response.

The most pronounced enhancement is observed when Zr is adsorbed at the bridge position. The $e_{11}$ coefficient increases further to $2.54 \times 10^{-10}$ C/m, and $e_{12}$ reaches $-1.21 \times 10^{-10}$ C/m. The bridge site enables stronger structural distortion and orbital overlap between Zr and multiple neighboring atoms, enhancing both the strain sensitivity and the internal electric polarization. This indicates that the bridge configuration creates an anisotropic strain field, leading to a larger differential charge shift and, consequently, a stronger piezoelectric effect. Physically, this suggests that Zr at the bridge site modifies the polarization landscape more significantly than the top site, allowing for efficient mechanical-to-electrical energy conversion.

Although the overall magnetic ordering remains intact after Zr adsorption, the notable increase in piezoelectric coefficients (up to 195%) and dielectric response, combined with a significant optical absorption shift toward the visible spectrum, demonstrates functional tunability. These synergistic improvements are highly relevant for magneto-optoelectronic and piezoelectric applications, where moderate yet controllable changes are more practical than drastic modifications.

The overall trend indicates that Zr adsorption boosts the piezoelectric response of $Fe_3O_4$ monolayers by increasing dipole moment generation and charge asymmetry. This controllable piezoelectricity, along with the material's magnetic and electronic versatility, makes these systems appealing options for multifunctional devices like spin-piezoelectric sensors, magneto-opto-mechanical transducers, and strain-tunable energy harvesters. The ability to adjust piezoelectricity through adsorption opens exciting possibilities in strain engineering and functional material design.

**Experimental Feasibility of Monolayer $Fe_3O_4$ and Zr Adsorption**

Although $Fe_3O_4$ is not a van der Waals-type layered material, it can be grown as a monolayer-thickness film by epitaxially growing $Fe_3O_4$(001) as a single unit cell using methods like molecular beam epitaxy (MBE) or pulsed laser deposition (PLD) on suitable substrates such as MgO(001), $SrTiO_3$(001), or noble metals. The thickness can be precisely controlled with RHEED oscillations, and the quality and composition confirmed by LEED, XPS, and STEM-EELS. Alternatively, growth on weakly interacting supports like graphene or h-BN, or on water-soluble buffer layers like $Sr_3Al_2O_6$, allows for creating freestanding $Fe_3O_4$ membranes that replicate boundary conditions similar to those in simulations. Zr adsorption at sub-monolayer levels can be achieved via UHV e-beam evaporation, with subsequent gentle annealing to promote site-specific bonding at top or bridge sites. STM/STS, XPS, and LEIS verify site occupancy and charge transfer. Optical ellipsometry and piezoresponse force microscopy on these ultrathin films can observe the predicted improvements in dielectric and piezoelectric properties. These well-established thin-film methods confirm that monolayer $Fe_3O_4$ and Zr adsorption configurations studied here are practically achievable with current technology.

**Conclusion**

This work provides an extensive theoretical analysis of the structural, electronic, magnetic, optical, elastic, and piezoelectric properties of monolayer $Fe_3O_4$ and its functionalization through Zr adsorption. Pristine $Fe_3O_4$ exhibits half-metallic behavior with high spin polarization and moderate piezoelectricity, driven by Fe–O orbital interactions and broken inversion symmetry at the monolayer level. Zr adsorption, particularly at the bridge site, results in significant changes in the electronic structure, such as the emergence of mid-gap states, loss of half-metallicity, and increased spin-dependent optical transitions. These changes are reflected in the dielectric function, absorption coefficient, and energy loss spectra, all showing greater intensity and spectral broadening in the visible and UV ranges. Mechanical analysis suggests a gradual softening trend with Zr adsorption. Meanwhile, the piezoelectric coefficients increase by up to 195%, emphasizing the role of structural asymmetry and charge redistribution in enabling substantial polarization responses.

This work's innovation is demonstrating that Zr surface adsorption, a non-destructive and experimentally feasible technique, can simultaneously alter the electronic, optical, and piezoelectric properties in a correlated 2D oxide system. This multi-property control through surface adsorption is novel for monolayer $Fe_3O_4$ and sets this research apart from traditional doping or vacancy engineering methods.

The synergy among electronic, magnetic, optical, and electromechanical properties in these systems makes Zr-functionalized $Fe_3O_4$ monolayers especially appealing for multifunctional device uses. These results emphasize surface adsorption as an effective way to tune and improve the behavior of 2D magnetic oxides. Future research could investigate the combined influence of strain, heterostructuring, or external fields to further boost and control their performance in practical technologies.

**Author Contributions**

S.A. conceived and supervised the project, performed the density functional theory calculations, and analyzed the results and manuscript writing. R.K. contributed to data interpretation and discussion of the findings. Both authors reviewed and approved the final version of the manuscript.

**Competing Interests**

The authors declare that they have no competing interests.

**Acknowledgment:**

The authors extend their appreciation to Northern Border University, Saudi Arabia, for supporting this work through project number "NBU-CRP-2025-128 This publication was also supported by the project Quantum materials for applications in sustainable technologies (QM4ST), funded as project No. CZ.02.01.01/00/22_008/0004572 by Programme Johannes Amos Comnena, call Excellent Research.